\documentclass[showpacs,prlprintnumbers,aps,prl,notitlepage,superscriptaddress,reprint]{revtex4-1}
\usepackage{graphicx}
\usepackage{amsmath,amssymb}

\usepackage{color}
\usepackage{float}
\usepackage{hyperref}

\definecolor{Zcolour}{rgb}{0.992, 0.588, 0.22}

\def\be{\begin{equation}}
\def\ee{\end{equation}}
\def\bea{\begin{eqnarray}}
\def\cH{{\cal H}}
\def\eea{\end{eqnarray}}

\begin{document}

\title{Doped Kondo chain - a heavy Luttinger liquid}
\author{Ilia Khait}
\affiliation{Physics Department, Technion, 32000 Haifa, Israel}
\author{Patrick Azaria}
\affiliation{Physics Department, Technion, 32000 Haifa, Israel}
\affiliation{Laboratoire de Physique Théorique des Liquides, Université Pierre et Marie Curie, 4 Place Jussieu, 75252 Paris, France}
\author{Claudius Hubig}
\affiliation{Arnold Sommerfeld Center for Theoretical Physics, LMU Munich, 80333 M\"unchen, Germany}

\author{Ulrich Schollw\"ock}
\affiliation{Arnold Sommerfeld Center for Theoretical Physics, LMU Munich, 80333 M\"unchen, Germany}

\author{Assa Auerbach}
\affiliation{Physics Department, Technion, 32000 Haifa, Israel}
\date{\today}

\begin{abstract}
The one dimensional $SU(2)$ Kondo Lattice model is studied by Density Matrix Renormalization Group away from half-filling. We find signatures of a Heavy Tomonaga-Luttinger Liquid (HTLL) phase, which describes one dimensional Heavy Fermions. We compute the wave-vector dependent charge and spin susceptibilities.  Our results establish divergent charge and spin correlations at the large Fermi surface $k_F+\pi/2$, and its harmonics.  We also find a signature of the hybridization gap at the small Fermi surface. We compare our $N\!=\!2$ results to the large-$N$ Slave Bosons mean field theory of the  $SU(N)$ Kondo Lattice model,
and find crucial effects of the RKKY interactions on the generation of renormalized effective mass for the $N\!=\!2$ model.
\end{abstract}

\maketitle

{\em Introduction \---- } The Kondo Lattice (KL)~\cite{Kasuya_1956, Doniach_1977, Hewson_1993} describes itinerant conduction electrons  Kondo-coupled to localized spins in each unit cell, by an $SU(2)$ symmetric magnetic interaction. The KL has been intensively studied as the microscopic model of heavy fermion (HF) metals in rare earth compounds. Nevertheless, much remains unknown about its low energy correlations.

The large-$N$ approximation of the $SU(N)$ model is given by Slave Bosons Mean Field Theory (SBMFT)~\cite{Coleman_1984, Read_1984, AssaLargeN, Millis_1987}. It describes a Fermi liquid of hybridized  conduction and valence electrons with a large Fermi surface. It includes conduction electrons and localized spins as itinerant fermions, in agreement with Luttinger's theorem~\cite{Luttinger_1960,Oshikawa_1997,Oshikawa_2000}. Signatures of the large Fermi surface were found numerically in one dimension~\cite{Tsunetsugu_1997}, and experimentally in HF compounds~\cite{Taillefer_1988, Aynajian_2012}.
The SBMFT predicts an exponentially large (in inverse Kondo coupling) effective mass $m^*$.
However, the validity of the SBMFT for the physical $N\!=\!2$ KL, has been severely challenged. Inter-site magnetic interactions, named RKKY~\cite{Ruderman_1954}, emerge at second order in the Kondo coupling, and in $1/N^2$~\cite{Houghton_1987}, but are neglected in the large-$N$ approximation. 
RKKY interactions may well destabilize the HF metal by magnetic ordering~\cite{Doniach_1977}, or at least change its low energy scales and correlations~\cite{Yang_2008}.
If a Fermi liquid phase is indeed found at weak coupling, the crucial question is:
{\it How does the effective mass $m^*$ depend on the Kondo coupling constant?} 

\begin{figure} [!hb]
    \centering
    \includegraphics[width=\linewidth]{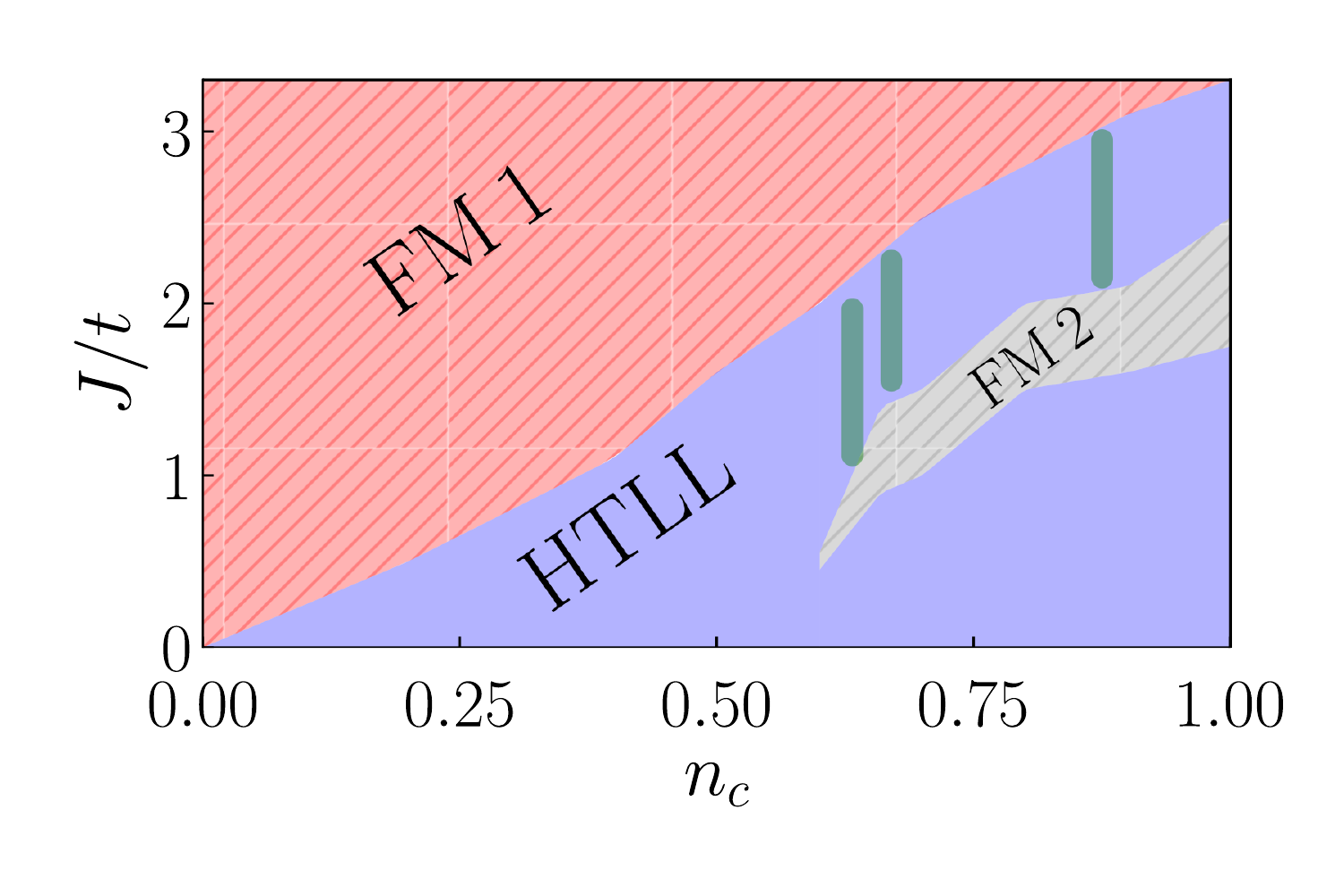}
    \caption{Schematic phase diagram of the Kondo lattice model in one dimension. $n_c$ is the conduction electron density, and $J/t$ is the dimensionless Kondo coupling constant.
    FM1 and FM2~\cite{Shibata_1996_Friedel,McCulloch_2002,Peters_2012}. are ferromagnetic regions. The paramagnetic phase (blue) is characterized as a ``Heavy Tomonaga-Luttinger liquid''. The DMRG calculations reported here are carried out at densities $n_c= 0.6, 0.67 \mbox{ and } 0.875$. }   
    \label{fig:PhaseDiagram}
\end{figure}

The one dimension (1D)  KL has been notoriously resistant to treatments by exact solutions, bosonization, quantum Monte Carlo and field theory, especially away from half-filling~\cite{Com1}.  Its phase diagram, see Fig.~\ref{fig:PhaseDiagram}, has been determined numerically~\cite{Tsunetsugu_1992,Troyer_1993,Tsunetsugu_1997} using exact diagonalization and Density Matrix Renormalization Group (DMRG)~\cite{White_DMRG_1992}. From the analysis of spin and charge density Friedel oscillations, and the large-$N$ approach~\cite{Shibata_1997} a Tomonaga-Luttinger Liquid~\cite{Haldane_1981} (TLL)
phase was detected, with a large Fermi surface~\cite{Oshikawa_1997} and a small Luttinger parameter (although some controversy remained~\cite{Xavier_2002,Xavier_2004}).  
The large-$N$ approach predicted an exponential mass renormalization, which has not yet been confirmed numerically.

In this paper we probe the TLL phase with new tools, in order to expose the low energy momentum and velocity scales.  
We use DMRG to 
compute the full  wave-vector dependent charge and spin susceptibilities. We obtain the {\em Heavy Tomonaga-Luttinger Liquid} (HTLL) properties, including the large Fermi wave-vector, Luttinger parameter, and both
spin and charge velocities. Away from half-filling ($n_c=1$ in Fig.~\ref{fig:PhaseDiagram}), the spin velocity, which is inversely proportional to the effective mass (e.g. $v_s \propto 1/m^*$), is found to scale as a {\em power law} of the Kondo coupling, in contrast to  the  large-$N$ prediction~\cite{Shibata_1997}. On the other hand, a charge gap is found at  twice the {\em small } (conduction electrons) Fermi wave-vector, 
which  is consistent with the {\em hybridization gap} predicted by the large-$N$ approach. 
We conclude  with a comparison between the N=2 results and the  SMBFT, which illuminates the effects of RKKY interactions on the  
low energy physics of the KL.

{\em Model \---- } We study the  SU(2) 1D KL model, 
\be
\mathcal{H}  = -t \sum_{i,s} c^\dagger_{i,s} c_{i+1,s}^{\phantom\dagger} + {\rm{H.c}} + J\sum_i \vec{S}_i \cdot \vec{s}_i,  
\label{eq:KondoHam}
\ee
$J$ is the Kondo interaction.  $c_{is}$ annihilates an electron at position $i=1\ldots L$ with $z$-spin $s$. $s^\alpha_i={1\over 2} \sum_{ss'} c_{is}^\dagger   \sigma^\alpha_{ss'} c_{i s' }^{\phantom\dagger}$,  and $S^\alpha$ are spin-half operators of the conduction electrons and localized f-spins, and $ \sigma^\alpha~,\alpha=x,y,z$, are Pauli matrices.   
The conduction electron number $N_c=n_c L $ defines the  ``small''  Fermi wave-vector  $k_F=  \pm \pi N_c/(2L)$. The ``large'' (Luttinger theorem) Fermi wave-vector is  $k_F^\star=k_F+{\pi\over 2}$.

The phase diagram in the $n_c$, $J/t$ plane, was determined previously by several groups~\cite{Sigrist_1992,Troyer_1993,Ueda_1993,Tsunetsugu_1997,Shibata_1999,Shibata_1996_Friedel,McCulloch_2002,Peters_2012,Com2},  as depicted in Fig.~\ref{fig:PhaseDiagram}.  
There are two ferromagnetic regions, FM1 and FM2, at strong and moderate Kondo couplings. In this paper, we  focused on establishing and characterizing the
paramagnetic regime as a Tomonaga-Luttinger Liquid (TLL) phase~\cite{Haldane_1981,GiamarchiBook}.  

{\em Observables  \---- }  
The uniform spin and charge susceptibilities are given by differentiating the  ground state energy  $E_0(L,N_c,M)$, where $M= \sum_i (S^z_i+s^z_i)$ is the conserved magnetization
\be 
\chi_s (L) =  {1\over L} \left({\partial^2  E_0 \over \partial M^2}\right)^{-1}, ~~~ \chi_c(L)  = {1\over L} \left( {\partial^2  E_0 \over \partial N_c^2}\right)^{-1}.
\ee
By the TLL theory, these are related to the spin and charge velocities by   
\be
 v_s = \chi_s^{-1}   /(2\pi) ,~~~~v_c \equiv  \chi_c^{-1}  K /(2\pi),
\label{vsvc}
\ee
where $K$ is the Luttinger parameter.
In a TLL, the spin and charge density Friedel oscillations near the boundaries are given to leading order by~\cite{Shibata_1996_Friedel,White_2002_Friedel}  
\be
\langle (S^z_i+ s^z_i) \rangle  = B_1 \cos(2k_F^\star x) \: {x_i}^{-K}, 
\label{eq:SpinFriedel}
\ee 
and
\be
\langle  \sum_s c^\dagger_{is}c_{is}^{\phantom\dagger}\rangle = A_1 \cos(2k_F^\star x_i) \: x_i^{-\frac{K+1}{2}} +  A_2 \cos(4k_F^\star x_i) \: x_i ^{-2K}. 
\label{eq:DenFriedel}
\ee
This allows us, in principle, to extract  $k_F^\star$  and  $K$. 
There are also sub-dominant contributions to 
Eqs.~\ref{eq:SpinFriedel} and~\ref{eq:DenFriedel} at other wave-vectors, which could allow us to extract more information about the TLL phase. However these are difficult to probe numerically, therefore we use a complementary approach, and compute the wave-vector dependent susceptibilities by adding source terms to the Hamiltonian $\cH=\cH_0+\cH'$, 
\be
\cH'= - h_q (S^z_q+ s^z_q) - \mu_q \rho_q ,\nonumber\\
 \ee
for an operator $O_i$, $O_q$ is the lattice sine transform.  The  susceptibilities are
 \be
 \chi_s(q) = -{1\over L}{\partial^2 E_0\over \partial h_q^2},~~~ \chi_c(q) = - {1\over L} { \partial^2 E_0 \over  \partial \mu_q^2} , 
\ee
where $h_q \mbox{ and }\mu_q$ must be taken to be numerically larger than finite size spin and charge gaps.

{\em Method \----}
We use open boundaries, with  U(1)~\cite{ITensor} and SU(2)~\cite{SyTen} DMRG~\cite{White_DMRG_1992,Schollwock_2011}. Lattice sizes were $L\le 192$. We retain  up to $5500$ states in the reduced density matrix.
We found that 28 sweeps were sufficient for good convergence.
The DMRG relative truncation error was less than $10^{-8}$.  

Friedel spin density oscillations were found at  twice the large Fermi  
wave-vector $2k_F^\star= 2k_F+\pi$, in agreement  with Luttinger's theorem.  
The signature of $2k_F^\star$ in the charge Friedel oscillations is too weak for detection~\cite{Sigrist_1992,Tsunetsugu_1992,Troyer_1993,Ueda_1993,Tsunetsugu_1997,Shibata_1999,Com2,Com3}.

We also determine the Luttinger parameter $K$ by measuring the power law singularities of the density operator $n_q$ at $2k_F^\star$ and $4k_F^\star$. To this end we use the expected scaling in the TLL theory
\be
n_{q} = - {\partial E_0\over \partial \mu_{q}} \sim \mu_{q}^{ \left(\Delta(q) \over 2-\Delta(q)\right)}, 
\label{Delta}
\ee
where $\Delta(q=2k_F^\star) = {K+1 \over 2}$, and $ \Delta (q=4k_F^\star)= 2K$ are the scaling dimensions of $n_q$ at the two wave-vectors $2k_F^\star$ and $4k_F^\star$. We find good agreement for the values of $K$ extracted from $\Delta$ and from the charge density Friedel oscillations in Eq.~(\ref{eq:DenFriedel}).

\begin{figure} [!ht]
    \centering
    \includegraphics[width=\linewidth]{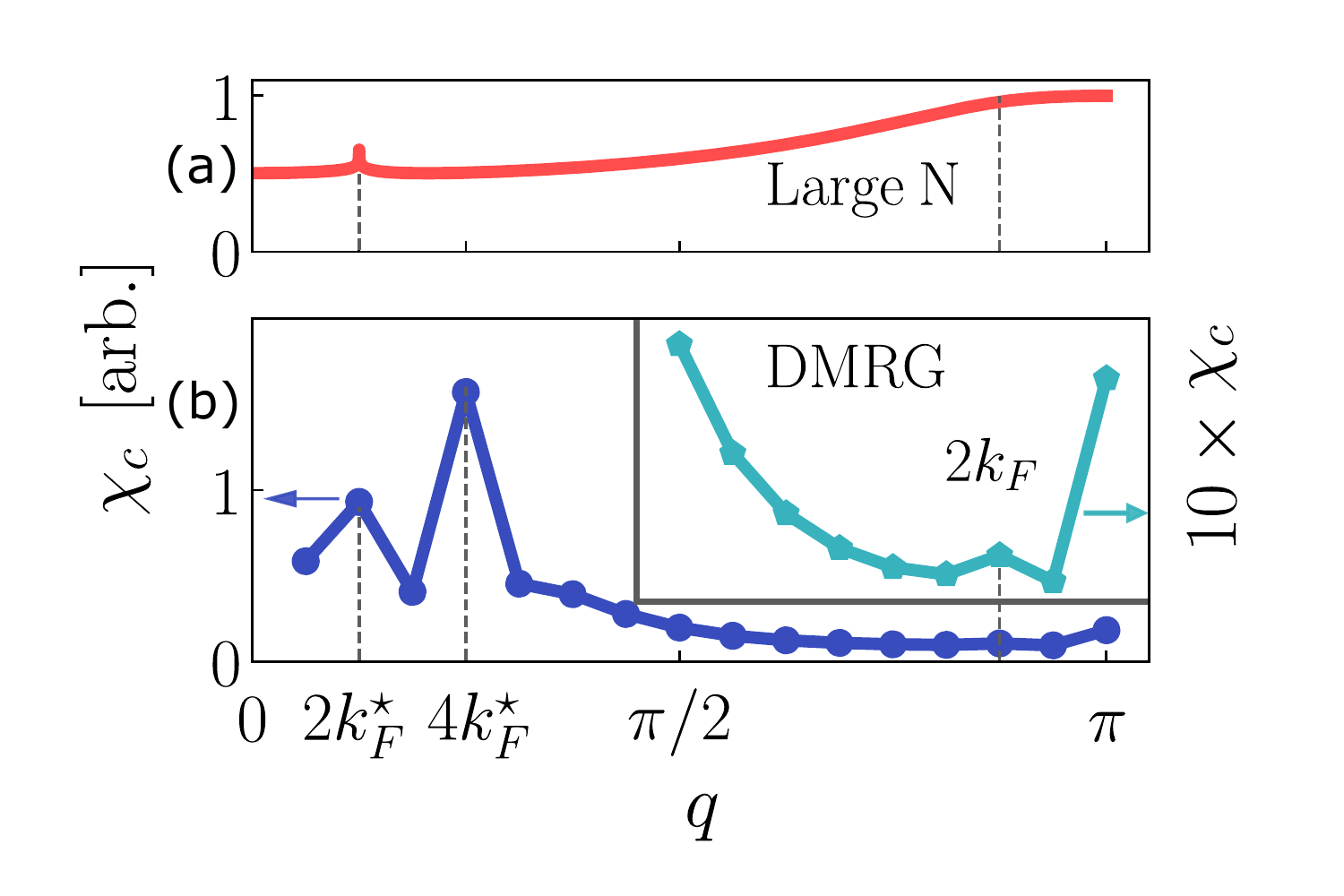}
    \caption{Charge susceptibility versus wave-vector, for $n_c=0.875, J/t=2.5$ (point (a) in Fig.~\ref{fig:PhaseDiagram}).   (a) SBMFT approximation. (b) DMRG calculation.  $k_F^\star$ denotes the large Fermi surface wave-vector. Finite  field scaling reveals divergent peaks at $2k_F^\star$ and $4k_F^\star$, as expected for a TLL. A weaker, non divergent, peak is found at twice the small Fermi surface $2k_F$,  which can be attributed to an inverse {\em hybridization gap}.}   
    \label{fig:ChargeSus}
\end{figure}

\begin{figure} [!ht]
	\centering
	\includegraphics[width=\linewidth]{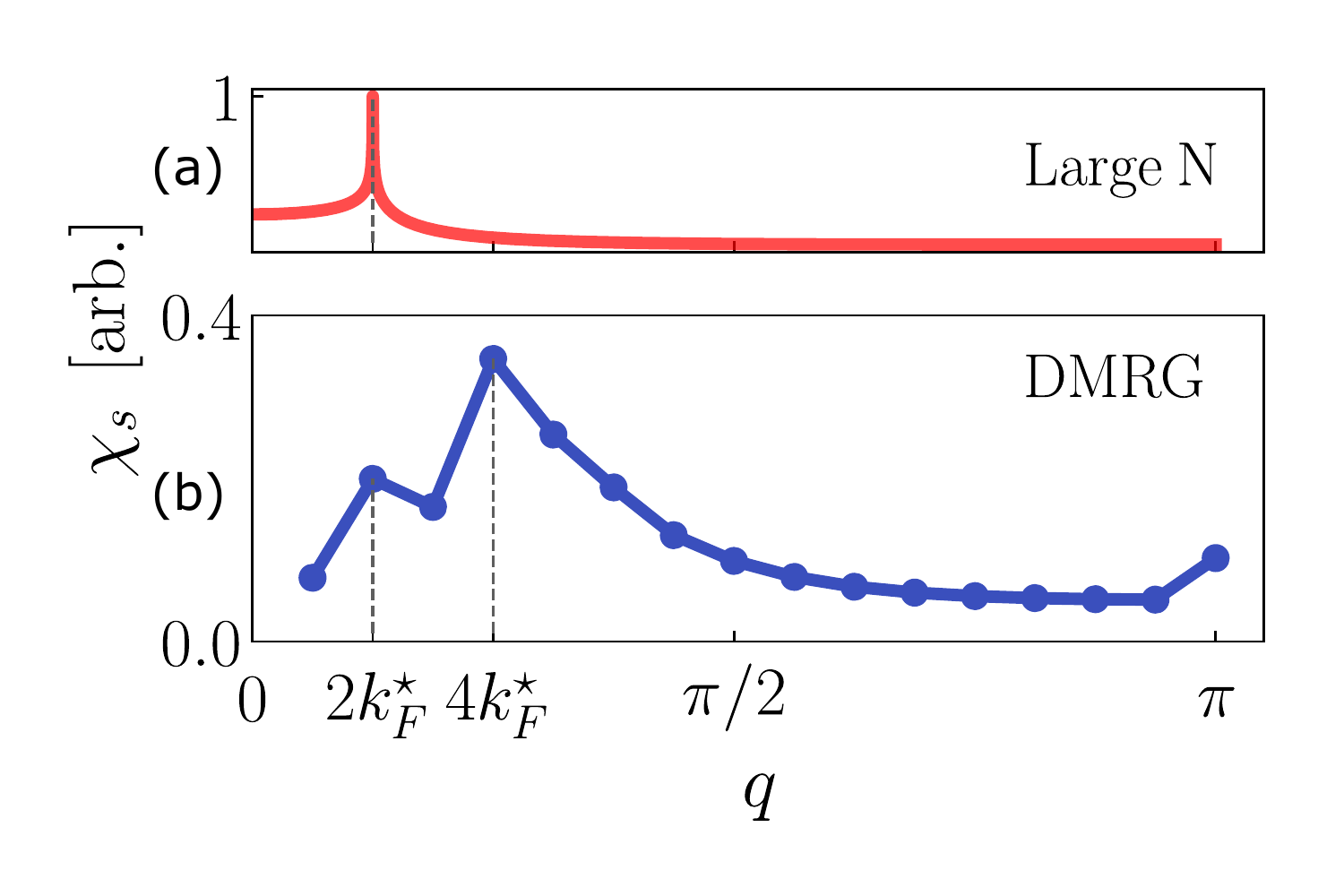}
	\caption{Spin susceptibility versus wave-vector, for $n_c=0.875, J/t=2.5$. (a) SBMFT approximation. (b) DMRG calculation.   Finite  field scaling reveals a divergent peak only at $q=2k_F^\star$.}   
	\label{fig:SpinSus}
\end{figure}

{\em Results \---- } We chose three fillings, $n_c$=0.6, 0.67, 0.875, with associated ranges of coupling constant $J/t$, within the paramagnetic region (shown in Fig.~\ref{fig:PhaseDiagram}).  
The lower values of $J/t$ were limited by  the rapid increase of ground state entanglement, which approached  the numerical limitations of our DMRG calculations.

The Friedel oscillations of the spin density are dominated by  wave-vector $2k_F^\star$. The charge density oscillations have $A_1 \ll A_2$, and since $K<0.33$, it is hard to resolve the sub-dominant $2k_F^\star$ oscillations~\cite{White_2002_Friedel}. 

Figures~\ref{fig:ChargeSus} and~\ref{fig:SpinSus} depict the bulk spin and charge susceptibilities respectively.
We note the detection of peaks at $2k_F^\star$ in {\em both} spin and charge sectors, which firmly confirm the TLL phase with  a large common Fermi wave-vector $k_F^\star$. 
The peak at $4k_F^\star$ marks the second harmonic of $2k_F^\star$, which is to be expected for a TLL with small values of $K$ (see Eq.~(\ref{eq:DenFriedel})).

\begin{figure} [!ht]
	\centering
	\includegraphics[width=\linewidth]{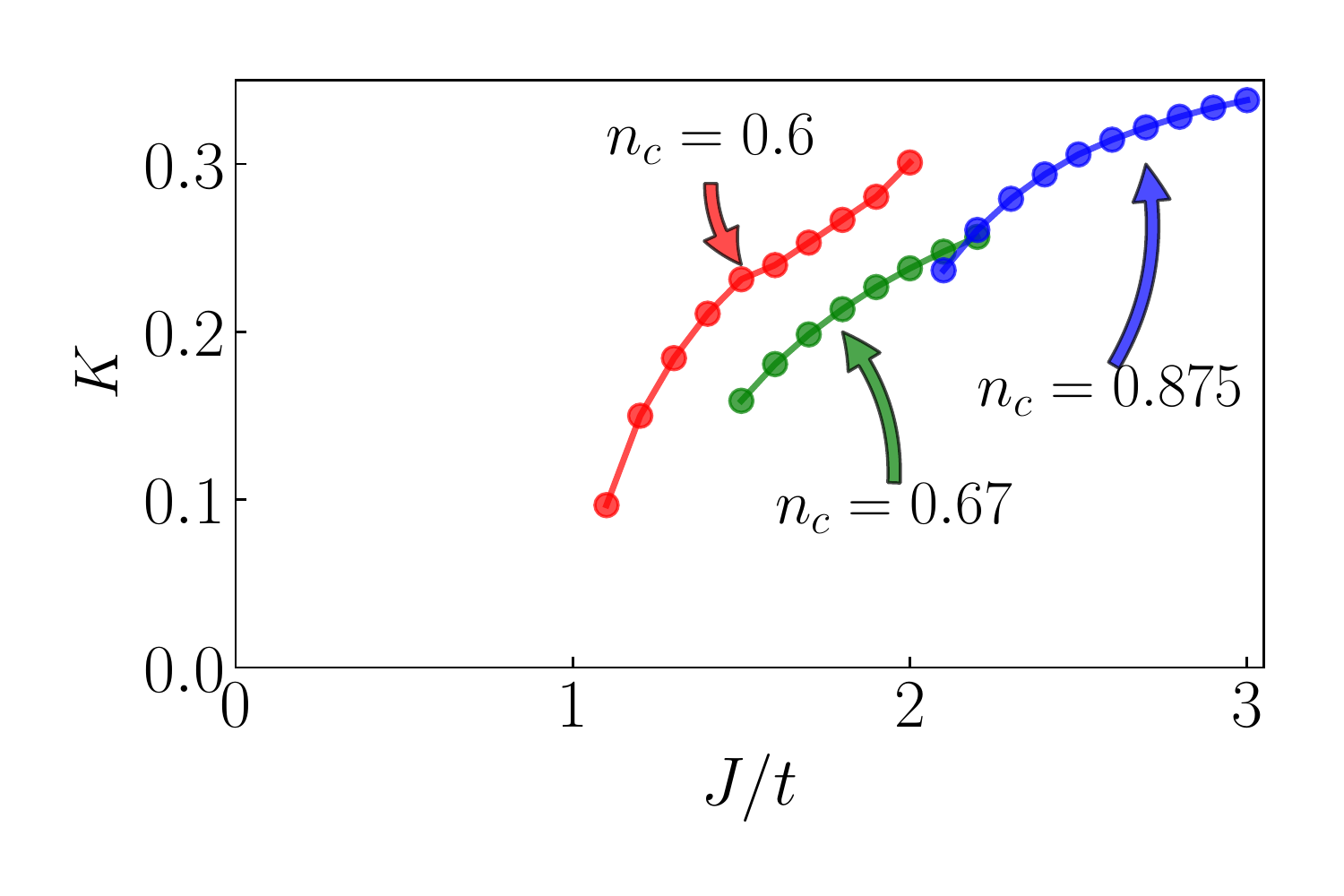}
	\caption{Luttinger parameter $K$ on the three lines shown in Fig.~\ref{fig:PhaseDiagram}.   The values of $K \lesssim 1/3 $ throughout the HTTL phase (see text) indicate effects of strong repulsive interactions.}     
	\label{figK}
\end{figure}

In Fig.~\ref{figK} the values of the Luttinger parameter $K(n_c,J/t)$ are depicted, we note a monotonic decrease as a function of $J/t$, signaling stronger repulsive interactions at weak coupling, as found earlier by Shibata et. al~\cite{Shibata_1997}. 

\begin{figure} [!ht]
	\centering
	\includegraphics[width=\linewidth]{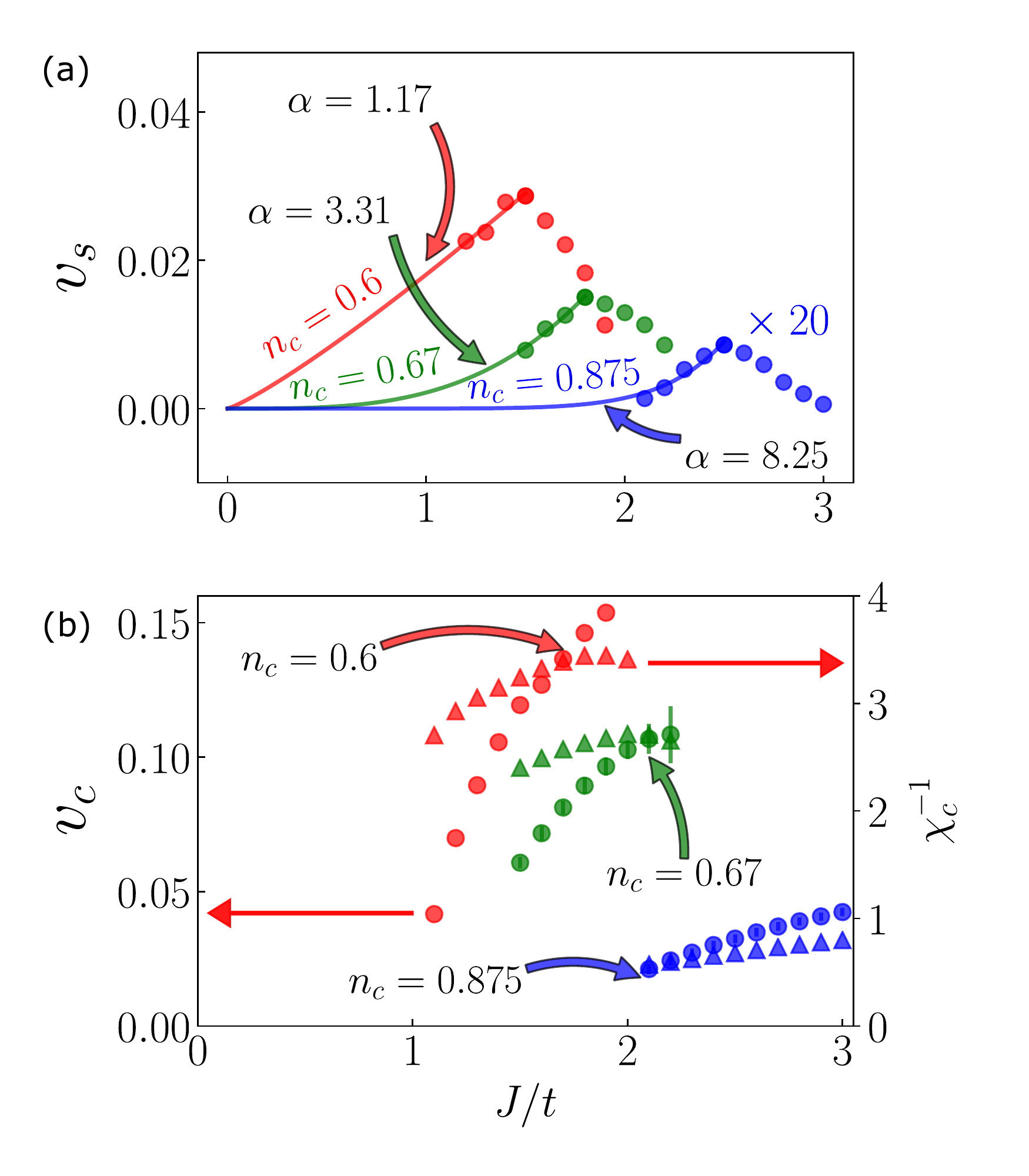}
	\caption{(a) Spin velocity as a function of $J/t$ at the different fillings. 
	 Solid  lines are power law fits $v_s = A(n_c) (J/t)^{\alpha(n_c) }$. We note a significant increase of $\alpha$ and a decrease of $A$ as $n_c \to 1$ (see discussion).
	 (b) Charge velocity (circles) and  inverse charge susceptibility (triangles) related by $\chi^{-1}_c\!=\!2\pi v_c/K$. Note the relatively weak $(J/t)$ dependence of $\chi_c$ (see discussion). }   
	\label{fig:vsvc}
\end{figure}

The spin and charge velocities of the TLL, as computed by Eq.~(\ref{vsvc}), depict qualitatively different behavior, as can be seen in  Fig (\ref{fig:vsvc}). The spin velocities, $v_s(J/t)$, are highly suppressed as the Kondo coupling $J/t$ is decreased. This ``heaviness'' is the hallmark of HTLL phase. We can fit the $v_s(J/t)$ data at weak coupling to power laws $v_s \propto (J/t)^{\alpha(n_c)}$, where $1.1 <  \alpha < 8.5$, which appears to increase sharply with $n_c$ toward the half-filled limit of $n_c=1$. 
In contrast, the charge velocity $v_c$ is that of the uncoupled conduction electrons Fermi velocity, and is given by $v_F^c= -2t\sin(k_F^\star)$. $v_c$ seems to be at least an order of magnitude larger, and varies moderately with $J/t$ and $n_c$.

\begin{figure} [!ht]
	\centering
	\includegraphics[width=\linewidth]{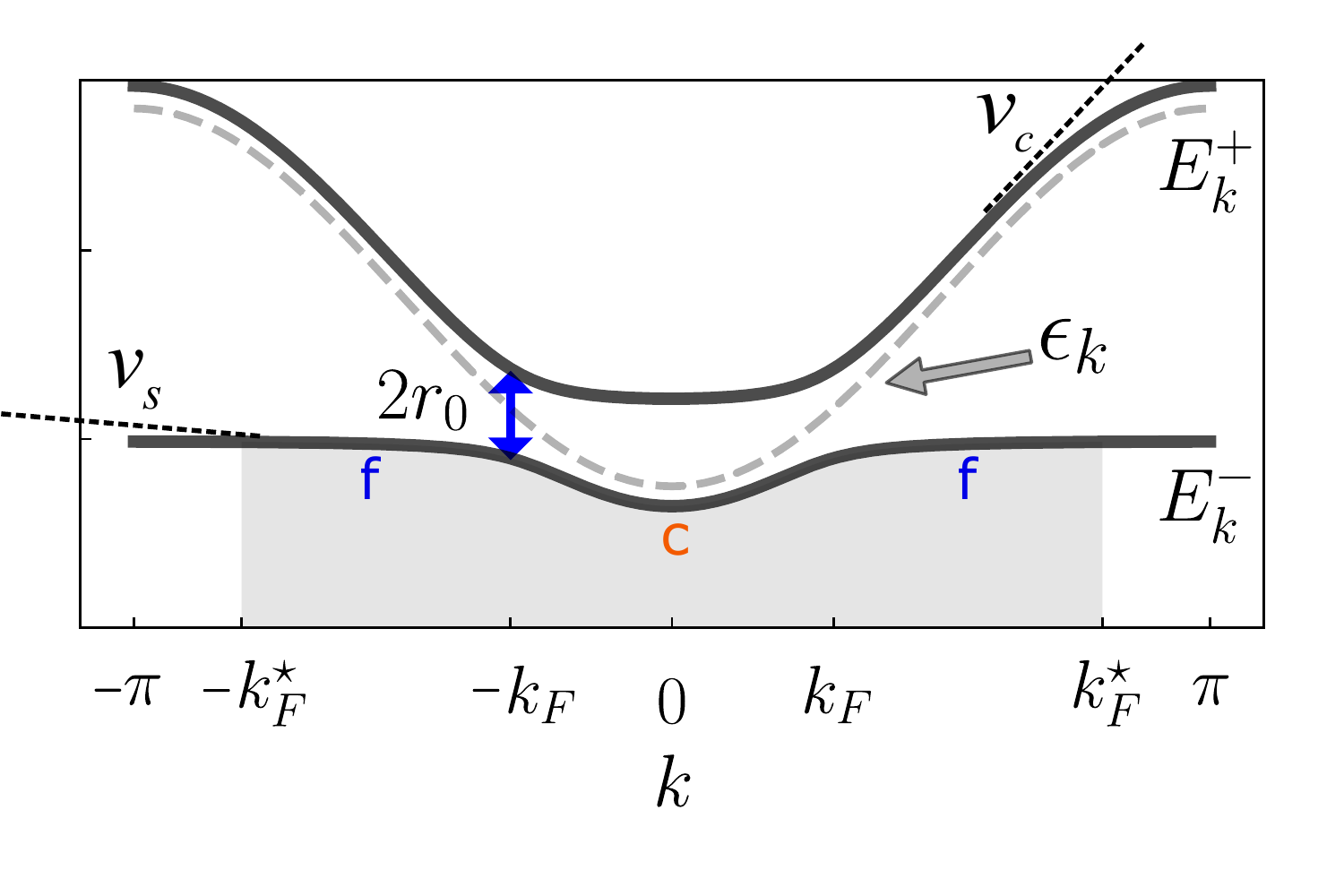}
	\caption{Band structure of the SBMFT Hamiltonian. $\epsilon_k$ is the tight binding kinetic energy, $E_k^\pm$ are the hybridized bands, $r_0$ is an hybridization variational parameter. Grey shaded area denotes the occupied part of the Fermi surface. Blue colored ``f'' and orange ``c'' denote the f-character and c-character parts of the band. Note the reduction in the velocity $v_s$ as compared to the Fermi velocity $v_c$.}   
	\label{fig:SBMFT}
\end{figure}

{\em Large-$N$ approximation \---- } 
The SBMFT as applied to Eq.~(\ref{eq:KondoHam}), yields a hybridized band structure,
\bea
\cH^{SBMFT} &=&  \sum_{ks}  \epsilon_k c^\dagger_{ks}c_{ks}^{\phantom\dagger} +  r_0  c^\dagger_{ks} f_{ks}+ {\rm H.c} + \epsilon_f f^\dagger_{ks}f_{ks}^{\phantom\dagger}\nonumber\\
&=& \sum_{ks}  E_k^\pm  \alpha^\dagger_{\pm,k,s} \alpha^{\phantom\dagger}_{\pm,k,s},
\eea
where $f_{k,s}$ represent the spins $S^\gamma_i = {1\over 2} \sum_{s,s'} \sum_{k,q} f^\dagger_{k,s}\sigma^\gamma_{ss'} f_{k+q,s'}^{\phantom\dagger}e^{-iqx_i}$ and $\sum_s f_{k,s}^\dagger f_{k,s}^{\phantom\dagger}=1$. The bare conduction electron band structure $\epsilon_k =  -2t\cos(k)$, and the hybridized bands $E_k^\pm$, are shown in Fig.~\ref{fig:SBMFT}, are
\be
E_k^\pm  ={ \epsilon_k + \epsilon_f\over 2}\pm \sqrt{\left( { \epsilon_k - \epsilon_f\over 2 }\right)^2 + r_0^2},
\ee
$\epsilon_f$ and $r_0$ are variational parameters, which depend on $J/t $ and $n_c$. Solving the mean field equations at weak coupling yields,
\be
r_0^2 = \left(  \epsilon_{k^\star_F} -  \epsilon_{k_F} \over 2 \rho_0\epsilon_{k^\star_F} \right) e^{1\over \rho_0 J},
\label{r02}
\ee
where $\rho_0 =  {1\over 4\pi t\sin(k_F^\star)}$ is the (single spin) conduction electron's density of states at the large Fermi wave-vector.
$r_0^{2}$ determines the Fermi velocity suppression as $v_F^\star = 2t r_0^2 \sin(k_F^\star)$, and $r_0$ yields the minimal hybridization gap at $k_F$. As shown in Fig.~\ref{fig:SBMFT}, $k_F$ marks a sharp crossover in the  character of the quasiparticles from $c$ to $f$ fermions.

{\em Discussion ---} 
We note that the SBMFT describes non-interacting fermions, (in essence a TLL with $K=1$).
Its spin and charge wave-vector dependent susceptibilities, are depicted for comparison with the DMRG in Figs.~\ref{fig:ChargeSus} and~\ref{fig:SpinSus}. SBMFT exhibits logarithmic singularity at $2k_F^\star$, which is weaker than the DMRG power laws analyzed by  Eq.~(\ref{Delta}) to yield $K<1$.
The DMRG exhibits an additional  peak in $\chi_c(q)$ at $4k_F^\star$, another consequence of $K<1$.

There are two  similar features of  the SBMFT picture and the DMRG results from which we can better understand the origin of the HTLL phase: (i) The relatively large difference between charge and spin susceptibilities In the SBMFT the character of the quasiparticles at $k_F^\star$ is largely that of $f$ fermions. Since $f$ electrons have zero charge susceptibility, its ratio is
\be
{\chi_s\over \chi_c} =  r_0^{-2}\propto e^{1\over \rho_0 J},
\ee
We find that for the KL, this ratio is indeed large. It differs from other TLL systems which derive from a single band model such as the Hubbard model~\cite{GiamarchiBook}.
Similarly to the SBMFT, $\chi_c^{-1}$ is a much weaker function of $J/t$ than the inverse spin susceptibility, as shown in Fig.~\ref{fig:vsvc}. (ii) The unique feature of the SBMFT is the existence of an intermediate energy scale, the minimal hybridization gap at the small Fermi wave-vector $E_{k_F}^+ - E_{k_F}^- \sim  2r_0$
(see Fig.~\ref{fig:SBMFT}. We find a signature of this gap at $2k_F$ in the charge susceptibility. This feature is observable in the SBMFT susceptibility at very weak coupling. Its existence is a strong indication that ``something is right'' about the hybridized band structure origin of the HTLL.
 
The most important  difference  between the large-$N$ SBMFT and the $N\!=\!2$ DMRG results, is the exponential versus power law dependence of $\chi_s(J/t)$. The SBMFT result indicates at a marginally relevant effect of the Kondo coupling at large-$N$. The power laws indicate the effects of a relevant perturbation, most probably of the emergent RKKY interactions, which are known to be logarithmically divergent at $2k_F$ in one dimension. In addition, in Fig.~\ref{fig:vsvc}, we see that the power law significantly increases toward $n_c \to 1$. It is numerically hard to distinguish between a power law and an exponential scaling for large $\alpha$, however we do see a clear power law dependence away from half-filling. The origin of this increase is still uncertain, but may be closely related to the phase transition at $n_c=1$ into a spin and charge gapped Kondo insulator~\cite{Tsunetsugu_1997}.

{\em Summary ---} Our DMRG results of the spin and charge wave-vector dependent susceptibilities,  establish the Heavy Tomonaga-Luttinger Liquid phase of the $N\!=\!2$ Kondo Lattice model at weak coupling and
away from half-filling.  We find  hybridized bands features in the numerical results. Analyzing the scaling of the Luttinger parameters with Kondo couplings and fillings sheds
interesting new light on understanding the relevancy of RKKY interactions, which are missing in the large-$N$ theory. In the future,  these insights may lead to finding a route to bosonize the one dimensional KL, and a way to understanding the role of RKKY interactions in higher dimensions.
 
{\em Acknowledgments -- } We thank Erez Berg, Sylvain Capponi, Anna Kesselman, and Daniel Podolsky for beneficial discussions. We are grateful to David Cohen for his extensive technical support. P.A. is grateful to the Physics Department of the Technion for their kind hospitality. A.A.  acknowledges United States - Israel Binational Science Foundation grant 2016168, and Israel Science Foundation grant 1111/16.  C.H. and U.S. acknowledge funding through the ExQM graduate school, the Bavarian Elite Network ExQM and the Nanosystems Initiative Munich.

\newpage

\end{document}


\renewcommand{\thesection}{S\arabic{section}}
	\renewcommand{\theequation}{S\arabic{equation}}
	\renewcommand{\thetable}{S\arabic{table}}
	\renewcommand{\thefigure}{S\arabic{figure}}
	\renewcommand{\thepage}{S\arabic{page}}
	\renewcommand{\vec}[1]{\boldsymbol{#1}}

\title{Supplemental Information for: Doped Kondo chain - a heavy Luttinger liquid}
\author{Ilia Khait}
\affiliation{Physics Department, Technion, 32000 Haifa, Israel}
\author{Patrick Azaria}
\affiliation{Physics Department, Technion, 32000 Haifa, Israel}
\affiliation{Laboratoire de Physique Théorique des Liquides, Université Pierre et Marie Curie, 4 Place Jussieu, 75252 Paris, France}
\author{Claudius Hubig}
\affiliation{Arnold Sommerfeld Center for Theoretical Physics, LMU Munich, 80333 M\"unchen, Germany}

\author{Ulrich Schollw\"ock}
\affiliation{Arnold Sommerfeld Center for Theoretical Physics, LMU Munich, 80333 M\"unchen, Germany}
\author{Assa Auerbach}
\affiliation{Physics Department, Technion, 32000 Haifa, Israel}
\date{\today}

\maketitle

\section{Susceptibility calculation}
We compute the wave-vector dependent charge ($\chi_c$) and spin ($\chi_s$) susceptibilities, as described in the main text. These are given by
\be
\rho(q,\mu_q) = \frac{e_0(q, \mu_q+\delta \mu_q) - e_0(q, \mu_q)}{\delta \mu_q}=\chi_c(q)\mu_q + \mathcal{O}(\mu_q^2),
\label{eq:DenSus}
\ee
and similarly, 
\be
m(q,h_q) = \frac{e_0(q, h_q+\delta h_q) - e_0(q, h_q)}{\delta h_q}=\chi_s(q)h_q + \mathcal{O}(h_q^2),
\label{eq:SpinSus}
\ee
$e_0(q,\mu_q)$ (or $e_0(q,h_q)$) denotes the ground state energy density in the thermodynamic limit in the presence of a charge (spin) density wave perturbation with wave-vector $q$ and amplitude $\mu_q$ ($h_q$).

\begin{figure} [!h]
    \centering
    \includegraphics[width=\linewidth]{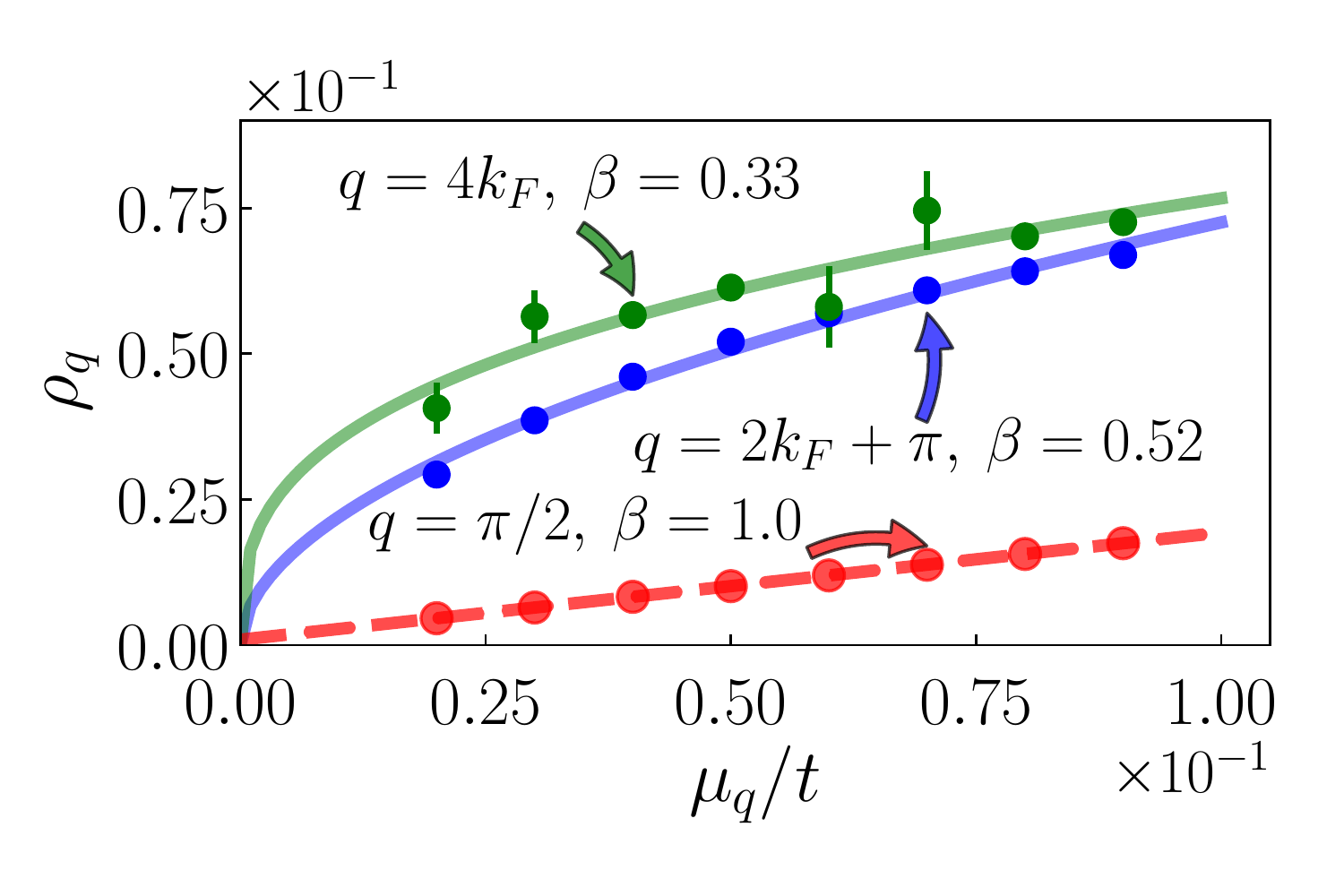}
    \caption{Finite field density response for $n_c=0.875$ and $J=2.5$, from which charge susceptibility is derived.}   
    \label{fig:DenRes}
\end{figure}

Finite field density response for $n_c=0.875$ and $J=2.5$ is depicted in Fig.~\ref{fig:DenRes}, from which charge susceptibility is derived. We show the response at three wave-vectors: $q=\pi/2$ (red), $q=2k_F^\star=2k_F+\pi$ (blue), and $q=4k_F^\star$ (green). All wave-vectors are fitted using a power law $\rho_q \sim \mu_q^\beta$. By using $\beta$ we find the wave-vectors for which the susceptibility diverges and extract the Luttinger charge interaction parameter. As an example, for $n_c=0.875$ and $J=2.5$ charge Friedel oscillations yield $K=0.3$. The density response scaling at $q=4k_F^\star$ yields $K=0.25$, and at $q=2k_F^\star$ we get $K=0.368$. Based on the trend of our data, if one could lower the amplitude of the applied fields one should see better agreement. 

In Figs.~\ref{fig:DenSus2} and~\ref{fig:SpinSus2} (see also main text) we plot the susceptibilities. For the wave-vectors in which the susceptibility diverges, we estimate the susceptibility by a linear extrapolation, i.e., the amplitude is the slope of the lowest applied fields. 

\section{Susceptibilities at a different filling}

\begin{figure} [!h]
    \centering
    \includegraphics[width=\linewidth]{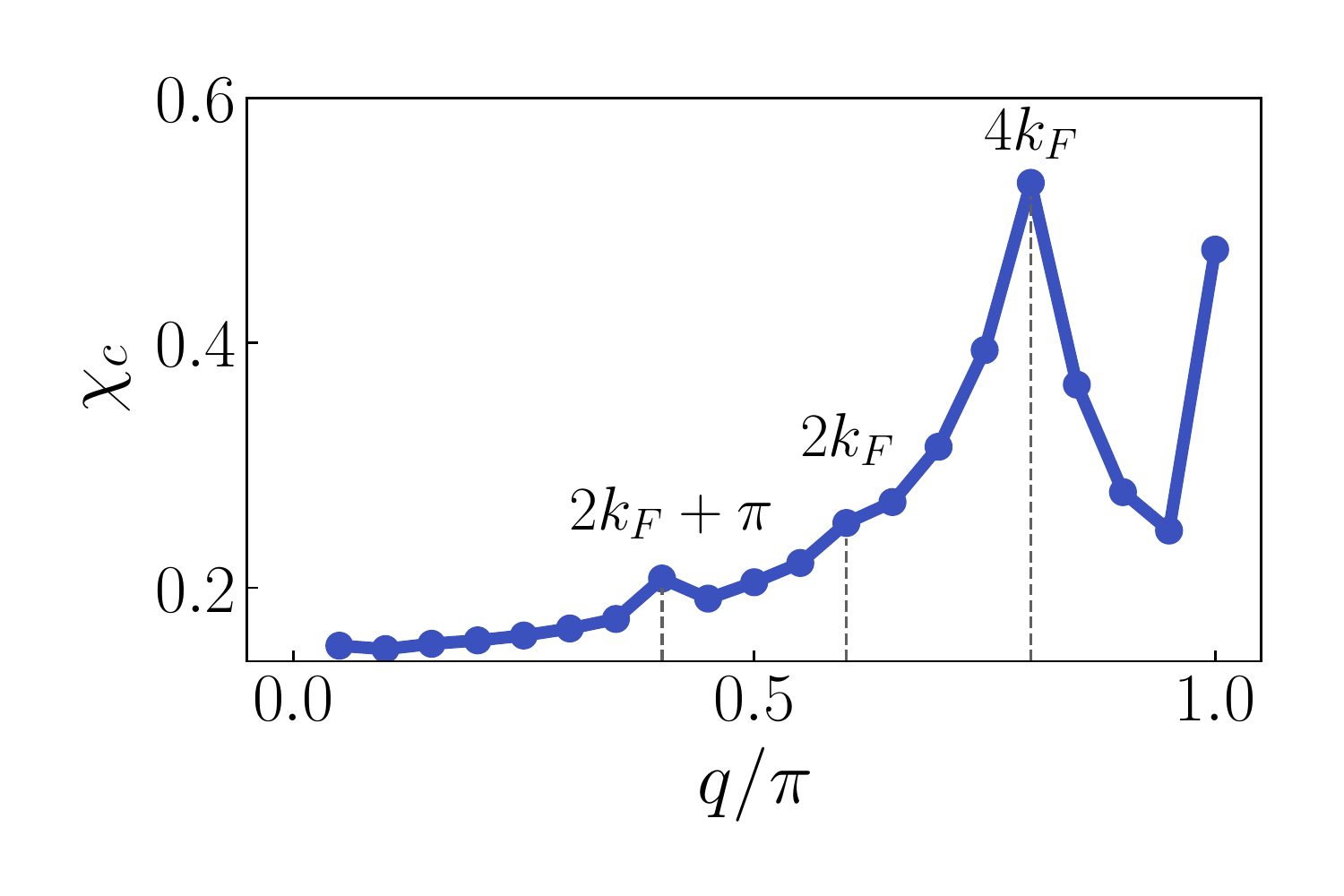}
    \caption{DMRG calculation of the charge susceptibility. Finite  field scaling reveals divergent peaks at $2k_F^\star = 2k_F+ \pi$ and $4k_F^*$, as expected for a TLL. A weaker, non divergent, peak is found at twice the small Fermi surface $2k_F$, which can be attributed to an inverse {\em hybridization gap}}  
    \label{fig:DenSus2}
\end{figure}

\begin{figure} [!h]
    \centering
    \includegraphics[width=\linewidth]{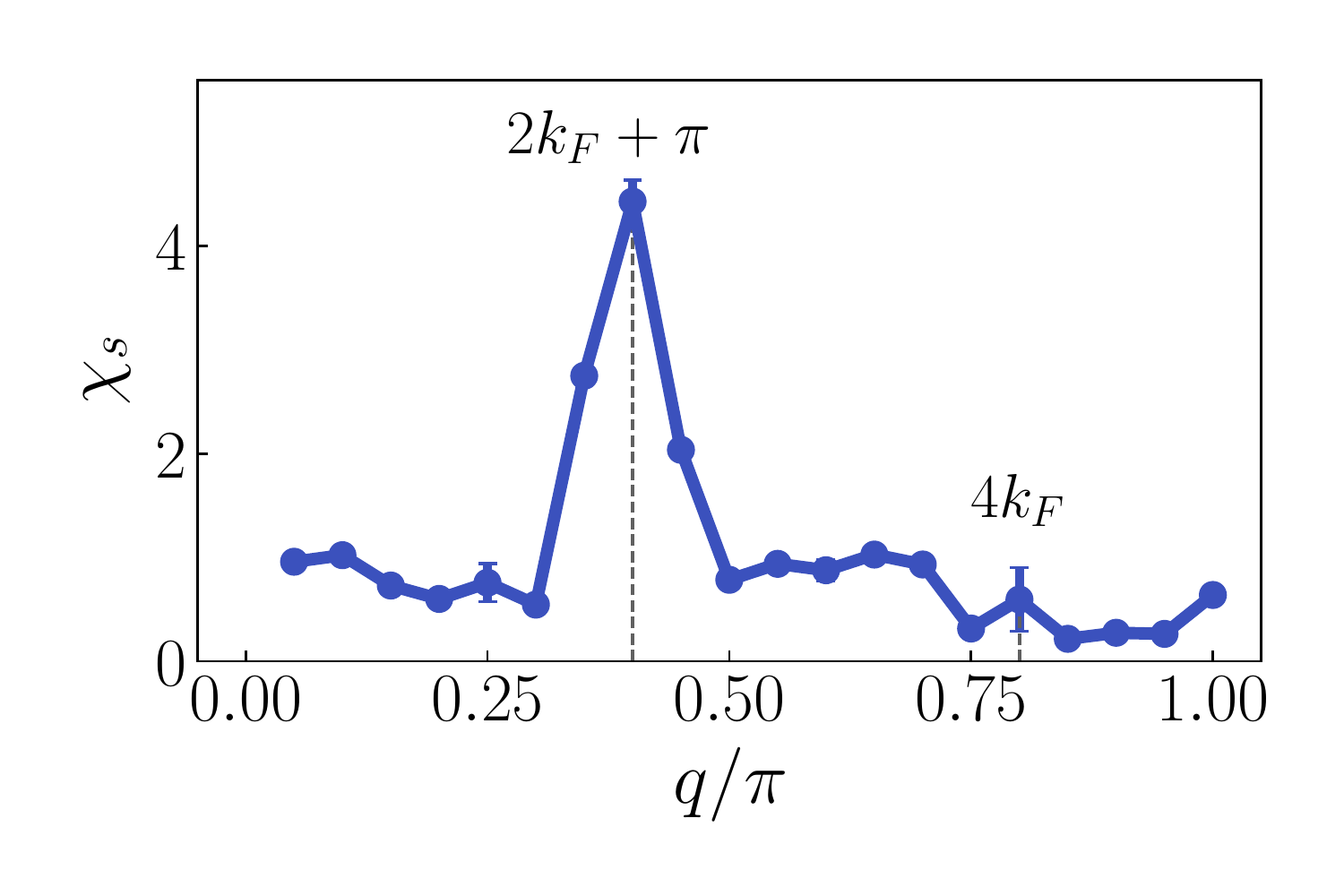}
    \caption{Spin susceptibility, finite field scaling reveals divergent peak only at $q=2k_F^\star$.}   
    \label{fig:SpinSus2}
\end{figure}

In addition to the density shown in the main text we also compute the susceptibilities for a conduction electron density of $n_c = 0.6$ and $J/t=1.5$. For both charge and spin susceptibilities the divergence occurs for the same wave-vectors, as seen in Figs.~\ref{fig:DenSus2} and~\ref{fig:SpinSus2}. The charge susceptibility has two diverging wave-vectors: $q=2k_F^\star \mbox{ and } 4k_F^\star$ and a feature at $q=2k_F$. As for the spin response, a divergence is seen at $q=2k_F^\star$. In general, due to small finite size spin gaps, and therefore extremely low spin velocities (see main text), convergence of the spin susceptibility is worse. Error bars are obtained using a comparison between positive and negative perturbation amplitudes.

\section{Friedel oscillations}

\begin{figure} [!h]
    \centering
    \includegraphics[width=\linewidth]{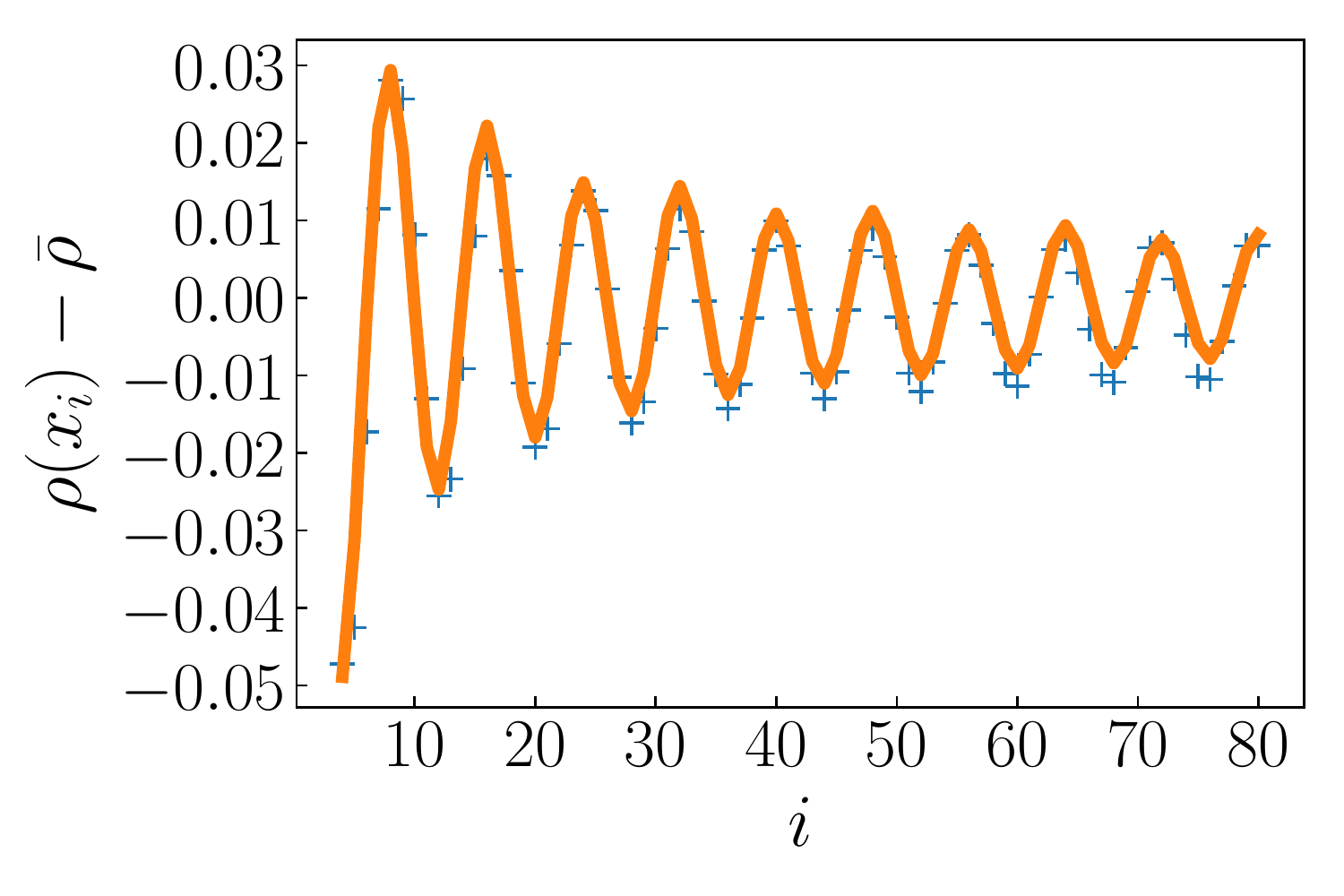}
    \caption{Friedel charge density oscillations for a system of $L=160$ sites. The orange line is a fit to Eq.~\ref{eq:DenFriedel}.}   
    \label{fig:FitFriedel}
\end{figure}

As it was discussed in the main text, we can compute the Luttinger parameter $K$, either by using the Friedel oscillations, may it be charge or spin, or by using the scaling relations of the response functions. Friedel oscillations offer two ways, we can either fit the entire function as given by
\be
\sigma_i = \langle (S^z_i+ s^z_i) \rangle  = B_1 \cos(2k_F^\star x) \:{x_i}^{-K},  
\label{eq:SpinFriedel}
\ee 
and
\bea
\rho_i = \langle  \sum_s c^\dagger_{is}c_{is}^{\phantom\dagger} \rangle &= A_1 \cos(2k_F^\star x_i) \: x_i^{-\frac{K+1}{2}} \nonumber \\  
&+  A_2 \cos(4k_F^\star x_i) \: x_i ^{-2K}, \label{eq:DenFriedel}
\eea
or we can fit the decaying envelope ($x^{-m}$, where $m$ is wave-vector dependent). If the Friedel oscillation data contains a single spatial frequency (i.e, only $q=4k_F$ oscillations), we find that it is more reliable to fit the entire function. This also allows for an estimate of the upper bound on the missing frequency's amplitude. An example is shown in Fig.~\ref{fig:FitFriedel}. We take system of length $L=160$, and perform a least square fit to Eq.~\ref{eq:DenFriedel}. According to the fit, $A_2=0.116 \pm 0.013$ while $A_1=0.009 \pm 0.005$, the Luttinger parameter is $K=0.31 \pm 0.02$. The large difference between the two amplitudes is a common feature to all fillings and interaction strengths. We believe that the $2k_F^\star$ mode is parametrically suppressed due to the strong repulsion in the model and therefore is hardly seen in charge Friedel oscillations.   
  
If there is more than one wave-vector present, we find it hard to filter the data without interfering with the extracted parameter values, therefore we fit to the envelope function, which induces a larger uncertainty. We repeat this procedure for each system size and extrapolate to the thermodynamic limit using $K(L) = K + a/L + b/L^2$.    

\begin{figure*} [!hb]
    \centering
    \includegraphics[width=\linewidth]{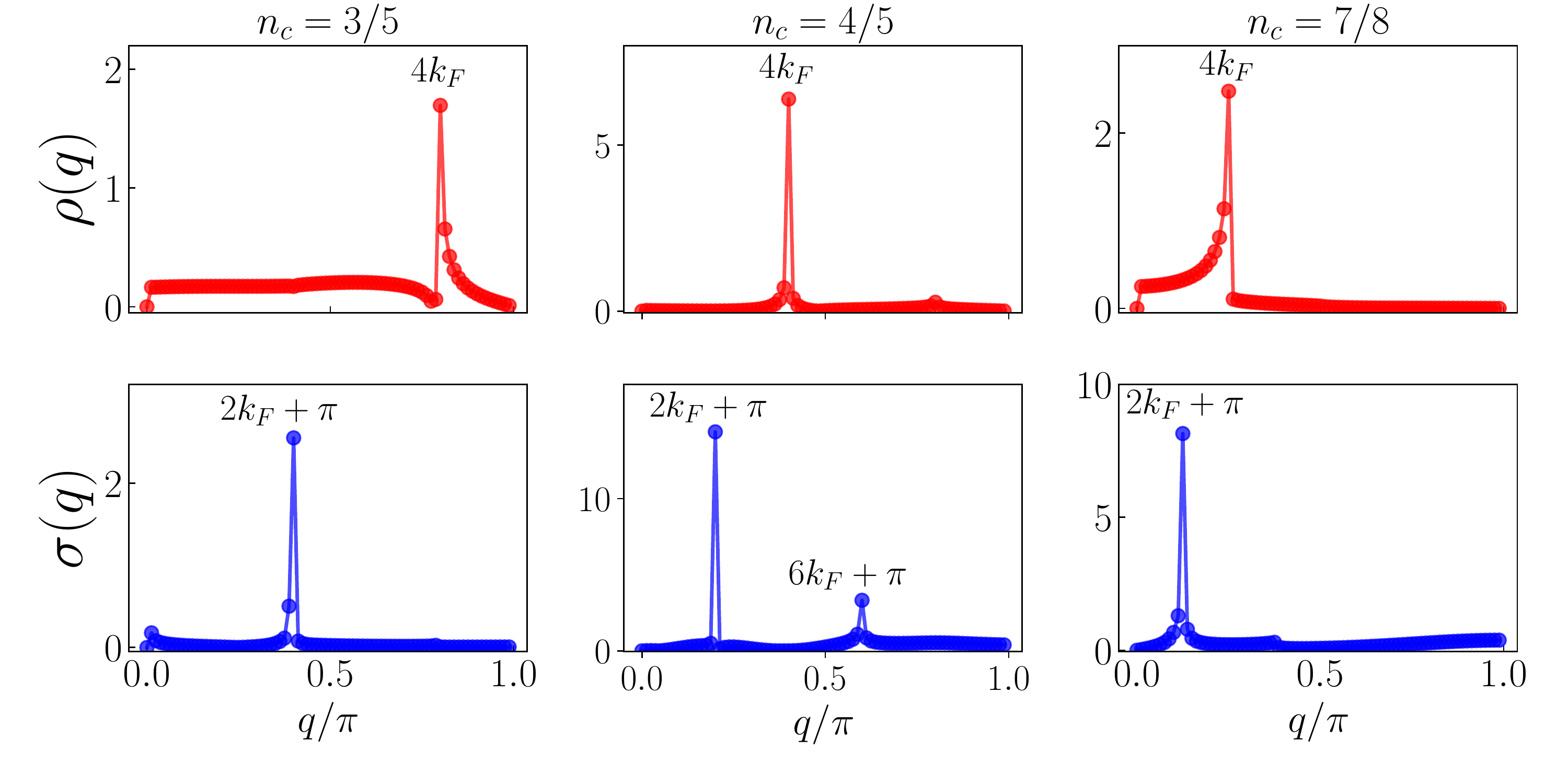}
    \caption{Fourier transform of charge density (red) and spin density (blue) Friedel oscillations. Various conduction electron densities ($n_c = 0.6,0.8,0.875$ from the leftmost panel to the right) show the same behavior. For the charge density Friedel oscillations, diverging (with respect to system size) peaks appear at $q=4k_F$, as seen in Fig.~\ref{fig:FO}. For the spin Friedel oscillations, the diverging peak appears at $q=2k_F + \pi$. At $n_c=4/5$, a higher harmonic can be seen at $q=6k_F + \pi$. We believe it is due to a $q=4k_F$ spin singlet mode which originates from the charge sector and a $q=2k_F + \pi$ from the spin sector.}   
    \label{fig:FO}
\end{figure*}

\section{Boundary perturbations}

Throughout the paper we used the following boundary perturbations in order to induce spin density Friedel oscillations. Opposite magnetic fields were applied at the boundaries as suggested by Shibata~\cite{Shibata_1996_Friedel}
\be
H_{\rm{SFO}}=h(s_z^1-S_z^1-s_z^N+S_z^N).
\ee
We varied the strength of the boundary field in the range $h=0.1 - 0.3t$, with no significant difference. 

\section{Analogy to the Hubbard model}

Friedel oscillations fail to reveal the large Fermi momentum in the Kondo lattice, this also occurs in the Hubbard model~\cite{White_2002_Friedel}. As one increases the on-site repulsion strength $U$ of the Hubbard model, one sees a suppression of the response at $q=2k_F$ with an increasing amplitude at $q=4k_F$, however {\it the response at $q=2k_F$ is present for any finite $U$}. For $U \rightarrow \infty$ the Luttinger parameter approaches $K=1/2$. Due to the scaling dimensions of the two terms in Eq.~\ref{eq:DenFriedel}, for $K>1/3$ the $q=2k_F^\star$ response is dominant, regardless of the magnitude of the non-universal amplitudes $A_1,A_2$. One can view this process as the loss of spin significance, as $U$ is increased, the fermions become effectively spinless. The fact that $K$ is bounded from below explains why the $q=2k_F$ amplitude never vanishes, and the spin-less limit explains why it is suppressed. 

In the Kondo lattice, the paramagnetic phase has a Luttinger parameter which does not exceed $K=1/3$. In this case, a priori $q=4k_F^\star$ should dominate. This is compatible with our findings. We suspect that if one goes to much larger systems, one could see a charge density Friedel oscillations response at $q=2k_F^\star$. The advantage of the susceptibility calculation lies in the fact that DMRG procedure is able to obtain ground state energies with high accuracy, and, while the non-universal amplitude of the susceptibility might be small (as for the Friedel oscillations), it is divided by the magnitude of the perturbation which is also small.

\newpage